\documentclass[10pt,twocolumn]{article}
\usepackage[margin=0.75in]{geometry}
\usepackage{graphicx}
\usepackage{lipsum} 
\usepackage{fancyhdr}
\usepackage{hyperref}
\usepackage{xcolor}
\usepackage{natbib}
\usepackage{float}
\usepackage{datetime} 

\hypersetup{
    colorlinks,
    citecolor=blue,
    filecolor=black,
    linkcolor=blue,
    urlcolor=blue
}

\title{User Guide to UVIT Data Reduction}
\author{Sujith Ranasinghe \& Denis Leahy} 
\date{} 

\newdateformat{monthyear}{\monthname[\THEMONTH] \THEYEAR}

\begin{document}

\maketitle

\section*{Abstract}

This document provides a user guide for reducing UVIT data using CCDLAB. While CCDLAB offers a straightforward data reduction work-flow, users may encounter certain challenges that require additional guidance. This guide provides instructions by addressing common issues related to key processing steps, including WCS solutions and VIS drift tracking.  

\section{Introduction}

The primary goal of this document is to provide detailed instructions on reducing Ultra Violet Imaging Telescope (UVIT) data, ensuring that users can effectively process and analyze their observations. While \cite{2021Postma} offers an extensive guide to UVIT data reduction using CCDLAB, including comprehensive explanations of the UVIT menu options, this document serves as a supplementary resource. It is designed to address specific challenges that users may encounter during the reduction process and provide practical solutions to enhance the efficiency and accuracy of data processing.

Data reduction using CCDLAB is generally straightforward, but certain steps may require additional clarification, especially when dealing with anomalies or unexpected issues. This guide aims to assist users in navigating these challenges by offering targeted troubleshooting strategies and step-by-step solutions.

In Section \ref{sec:datared}, the fundamental procedure for data reduction is outlined, covering essential steps to process UVIT data efficiently. Although determining the World Coordinate System (WCS) solutions is an integral part of this process, users may face difficulties that require specialized attention. Therefore, WCS troubleshooting is discussed separately in Section \ref{Sec:WCS}, providing guidance on resolving common errors and improving the accuracy of image alignment. Additionally, Section \ref{sec:VIS} provides concise instructions for handling VIS drift tracking issues, which may arise due to faint sources or tracking inconsistencies.

To further support users, high-quality video tutorials by Joe Postma are available for each of these sections. These tutorials offer visual demonstrations and additional explanations to supplement the written instructions. The URLs for these videos are provided at the end of each relevant section.
 
\section{Level 1 UVIT Data Reduction} \label{sec:datared}

In this section, Necessary steps for UVIT data reduction using CCDLAB is described. CCDLAB could be downloaded at \url{https://github.com/user29A/CCDLAB}. 

Navigate to the UVIT Menu and select ``Extract gz or zip Archives" to process the downloaded UVIT Level 1 data. Ensure that 7-Zip is installed, as it is required to extract the compressed files. During this step, the program automatically extracts the data and proceeds with preliminary processing, upto orbit-wise registration.

\begin{figure}[h]
  \includegraphics[width=\columnwidth]{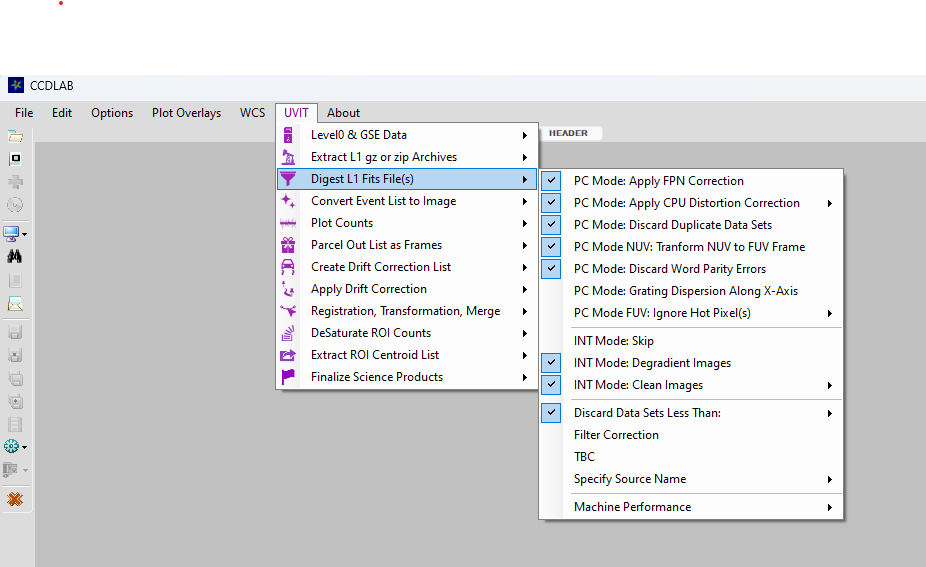}
  \caption{Level 1 FITS digestion.}
  \label{fig:1}
\end{figure}

\begin{figure}[h]
  \includegraphics[width=\columnwidth]{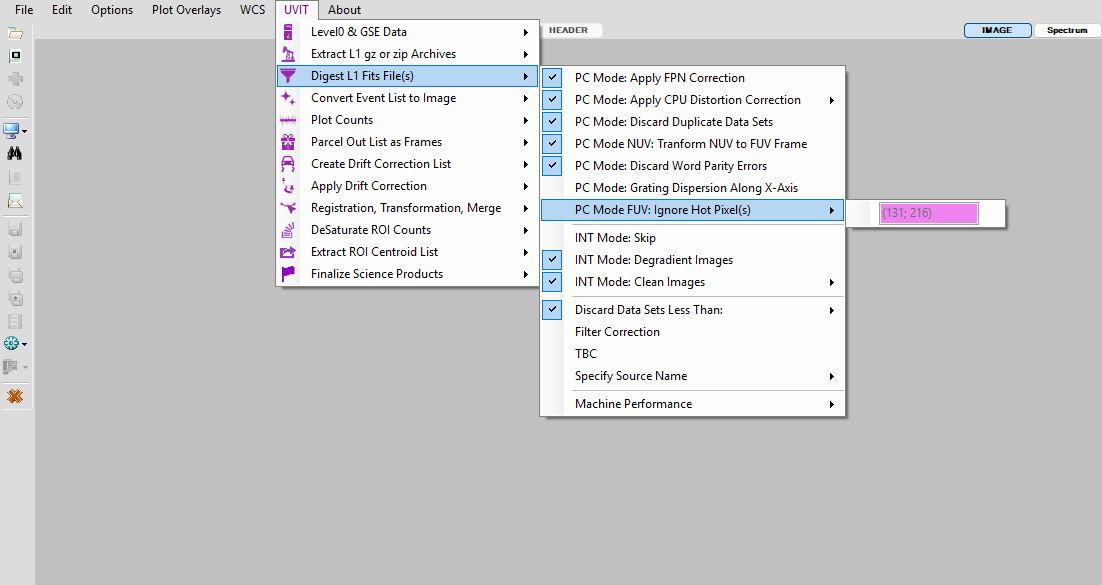}
  \caption{Ignore hot pixel option.}
  \label{fig:2}
\end{figure}

\begin{figure}[h]
  \includegraphics[width=\columnwidth]{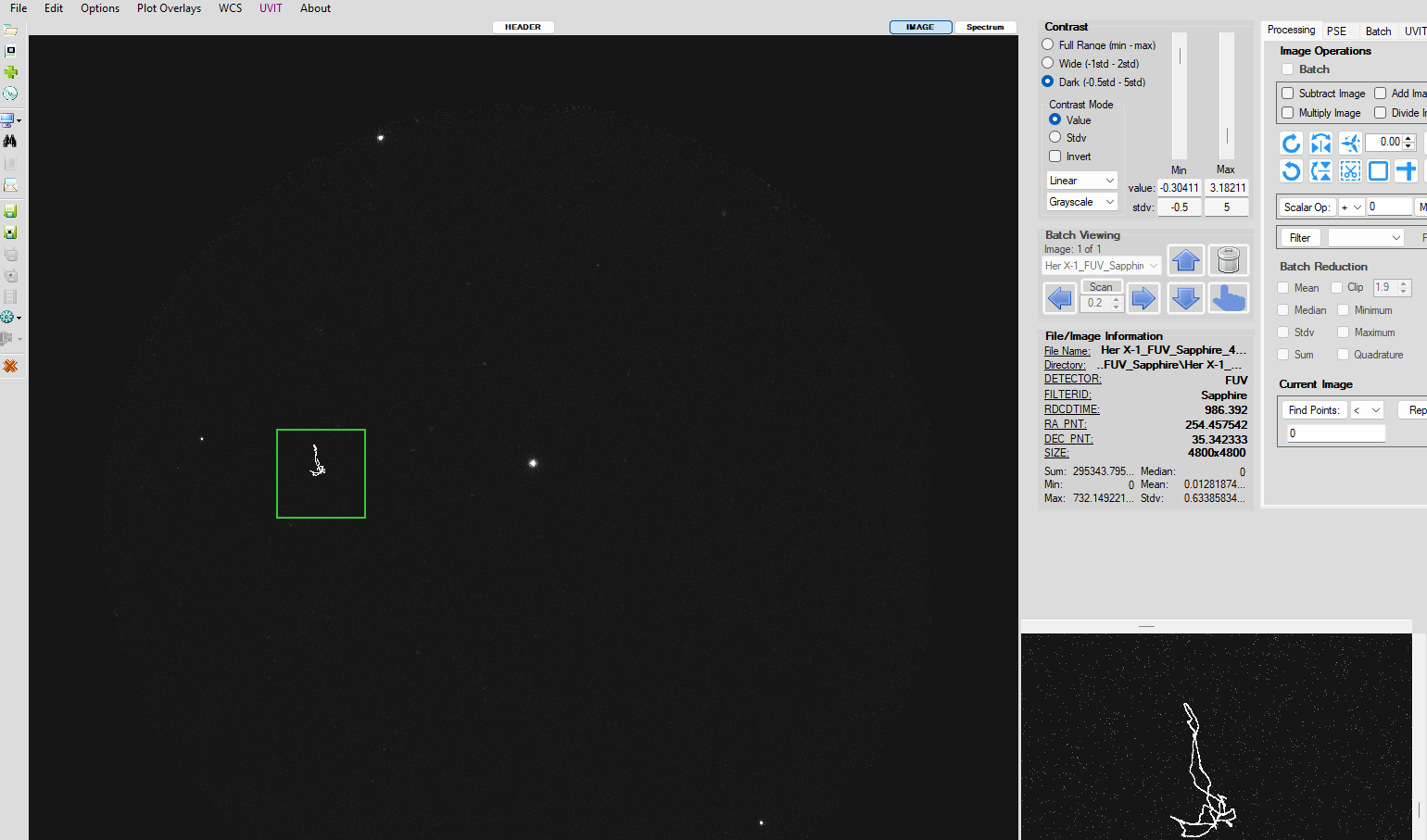}
  \includegraphics[width=\columnwidth]{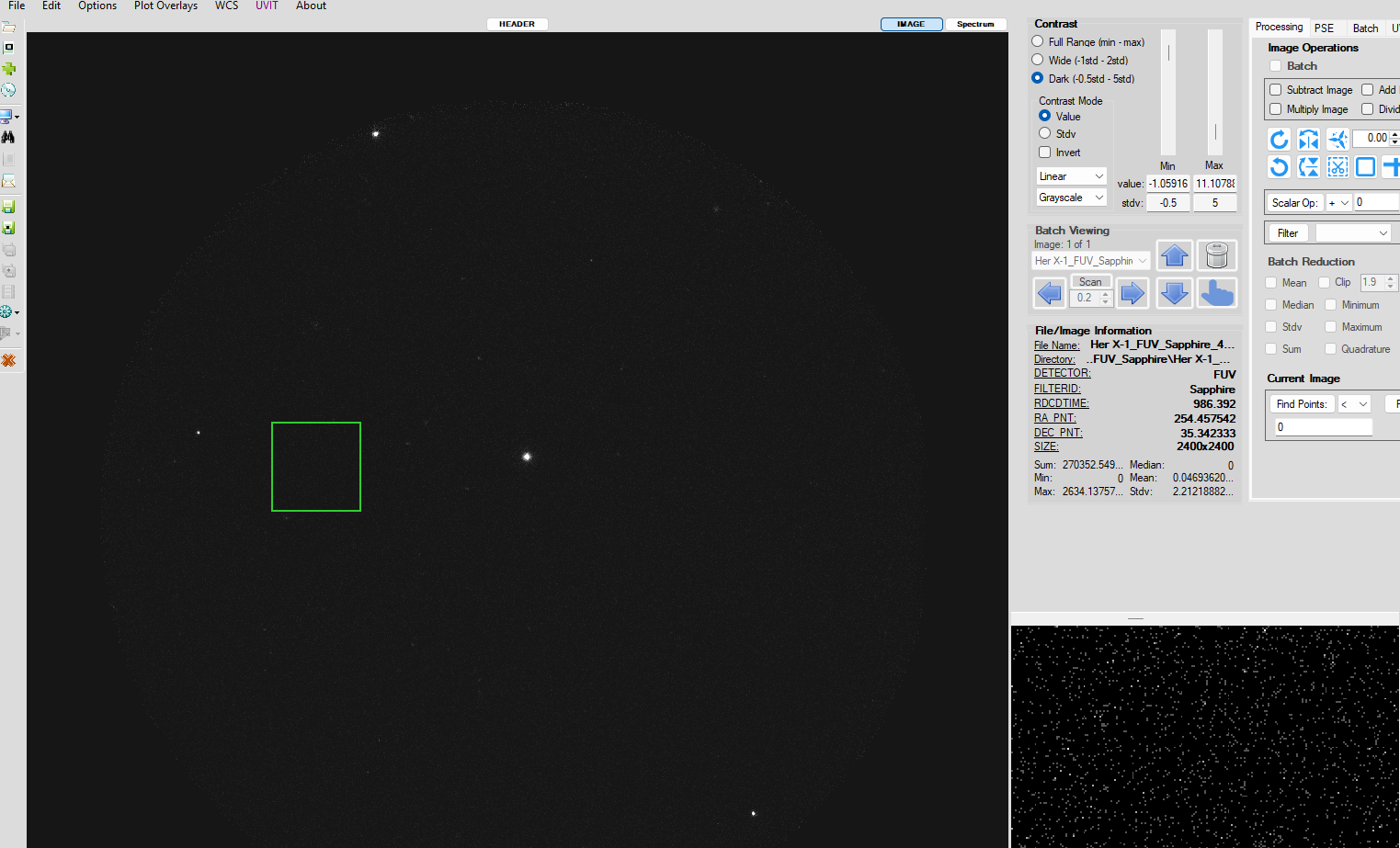}
  \caption{Example of a hot pixel in Image (Top) and hot pixel ignored image (Bottom).}
  \label{fig:3}
\end{figure}

 Figure \ref{fig:1} shows the current Digest L1 Fits File(s) menu with a few additional options (compared to \cite{2021Postma} Figure 4). Unlike the older version of CCDLAB, There is an option here to ignore hot pixels (Figure \ref{fig:2}). It should be noted that the presence of a hot pixel is seen at the beginning of the orbit-wise registration. Therefore, the data needs to be re-processed with the hot pixel option checked. Figure \ref{fig:3} shows an example of an image with a hot pixel (Top) and an image with the hot pixel removed (Bottom).  
 
Once the Level 1 data is extracted, the centroid lists (FUV and NUV) and Int mode visible tracking data are processed, followed by VIS drift tracking. The VIS drift tracking is automated and runs as part of the extraction. However, in some cases with faint sources, the VIS drift series may not work as expected, resulting in an error. In such cases, point sources need to be selected manually. Section \ref{sec:VIS} provides a description of how to do this.

Once the extraction is complete, the next step is registration, as the individual orbital images may have some rotation. To merge the images, registration is performed using the UVIT menu: ``General Registration" under the``Registration, Transformation, Merge". Then, navigate to the appropriate directory.

In this process, a few bright point sources are selected by left-clicking on them (marked with a red box). To move to the next image, right-click. The first selected point serves as the anchor point. If the red boxes around the point sources are not aligned in subsequent images, the anchor box can be adjusted to the correct location, and the other boxes can be moved or rotated until all sources are properly aligned. If the point sources and red boxes are not properly aligned, the registration step can be re-done. Since this is an iterative process, registration can be done as needed.  

Next, the images could be merged, using the UVIT menu: ``Merge Centroid List" under the ``Registration, Transformation, Merge" navigating to the appropriate directory. As an option, the source profiles of the merged image can be optimized using the UVIT menu: ``Create Drift Correction List": ``From PC Mode List": ``Optimize Point Source ROI".

An on-line tutorial can be found at \url{https://www.youtube.com/watch?v=QyIfjwN4DK4}, which gives step-by-step instructions on reducing UVIT Data. The next step is to find the World Coordinate Solutions (WCS) for the images and is described in Section \ref{Sec:WCS}.

\section{World Coordinate Solutions} \label{Sec:WCS}

\begin{figure}[h]
  \includegraphics[width=\columnwidth]{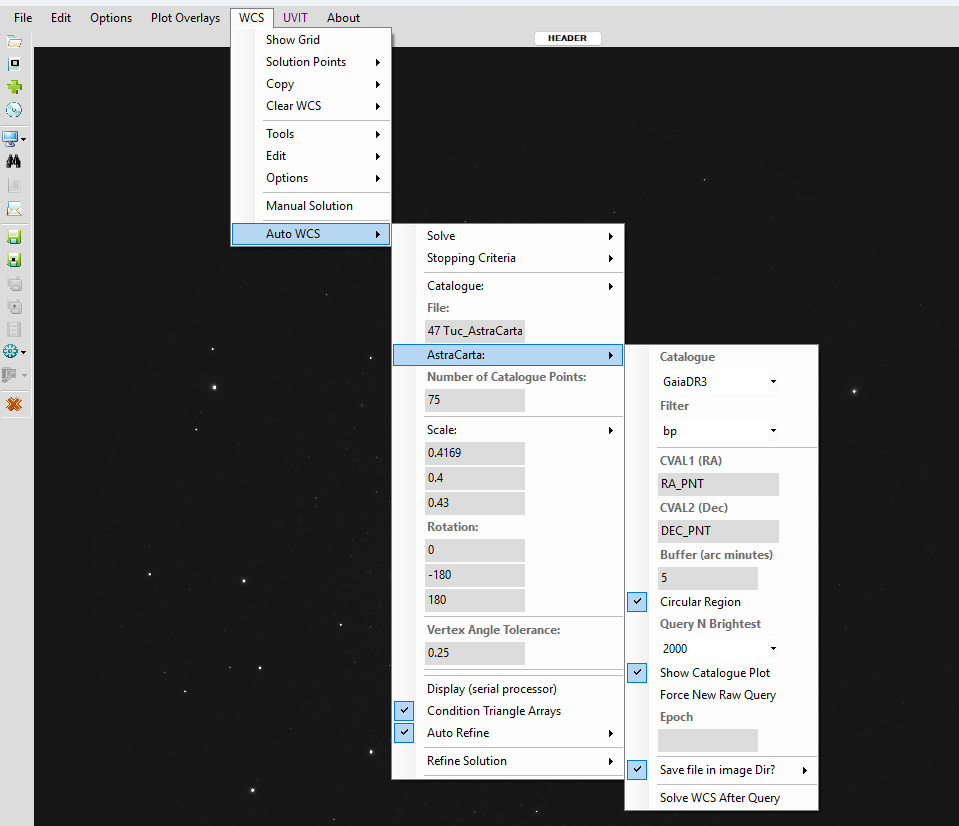}
  \caption{CCD World coordinate solver.}
  \label{fig:4}
\end{figure}

Finding the World Coordinate System (WCS) solution for the merged image is the next step in the process. Initially, the newly installed CCDLAB contains default values that are not accurate. Figure \ref{fig:4} provides the necessary values that should be entered.

The older version of CCDLAB used AstroQuery with the Gaia DR2 catalogue, whereas the newer version uses the Gaia DR3 catalogue. Additionally, the updated version allows users to specify the filter (bp, g, rp) based on the pass bands published by \cite{2010Jordi}.

The CVAL1 (RA) and CVAL2 (Dec) should be selected from the drop-down menu as RA\_PNT and DEC\_PNT, respectively. Once the correct values are entered, double-click on ``AstraCarta" to download the necessary catalogue, which will be saved as a FITS file. Then, by clicking ``Solve," the software will proceed to compute the WCS solution.

\begin{figure}[h]
  \includegraphics[width=\columnwidth]{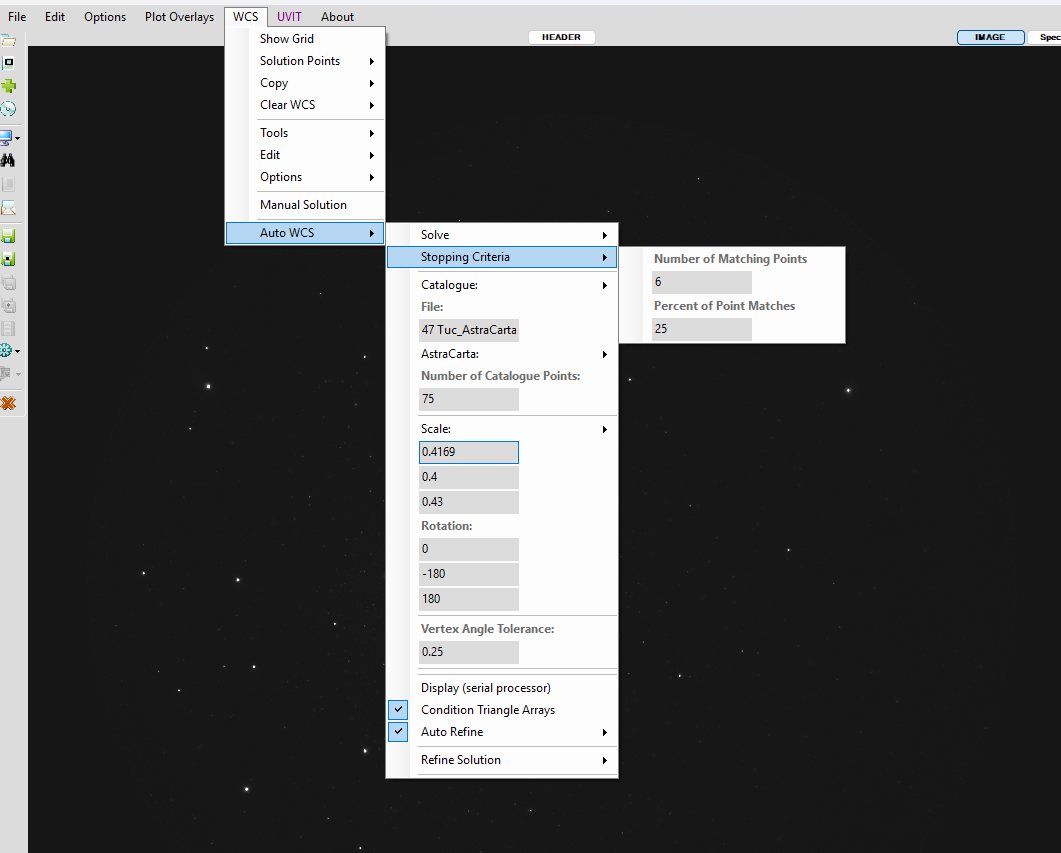}
  \caption{CCD World coordinate solver: Stopping criteria.}
  \label{fig:5}
\end{figure}

With the stopping criteria set to a default of 6 and the number of catalogue points set to 75 (see Figure \ref{fig:5}), the WCS is generally found. However, in some cases, a pop-up box may appear stating ``No solution." If this occurs, increasing the number of catalog points may help. It should be noted that increasing the number of catalogue points significantly increases the processing time. Once a solution is found, the software refines it using a certain number of sources.

In some cases, the WCS may not be accurate because multiple solutions are possible. This often occurs when the solution is refined using a number of sources equal to or fewer than the stopping criteria. To avoid this, increase both the stopping criteria sources and the number of catalog points. For example, increase the stopping criteria by one and the catalog points by 25. The software should be run multiple times to ensure that only a single solution is obtained.

Once the WCS is found, the image may be rotated. To de-rotate the image, navigate to UVIT Menu → ``De-rotate Loaded Images via WCS" under the ``Registration, Transformation, Merge" option. The blue-red lines indicate declination (blue = north, red = south).

If all else fails, WCS can be obtained by manually entering coordinates for known point sources. However, some challenges may arise, such as identifying specific point sources in a crowded field and de-rotation issues. Therefore, this method should be considered a last resort.

The final step is to finalize the science product, which can be found in the UVIT menu as an option. Double-clicking on this option removes all intermediate processing folders and files, leaving only the final image(s), exposure map(s), and the archived original Level 1 zip file.

In some cases, artifacts may appear in the point sources of the final image. A tutorial on resolving this issue is available at: \url{https://www.youtube.com/watch?v=jPk-gheWeCg}. The WCS tutorial video can be accessed here: \url{https://www.youtube.com/watch?v=4_48yRcN3nc}.

\begin{figure}[h]
  \includegraphics[width=\columnwidth]{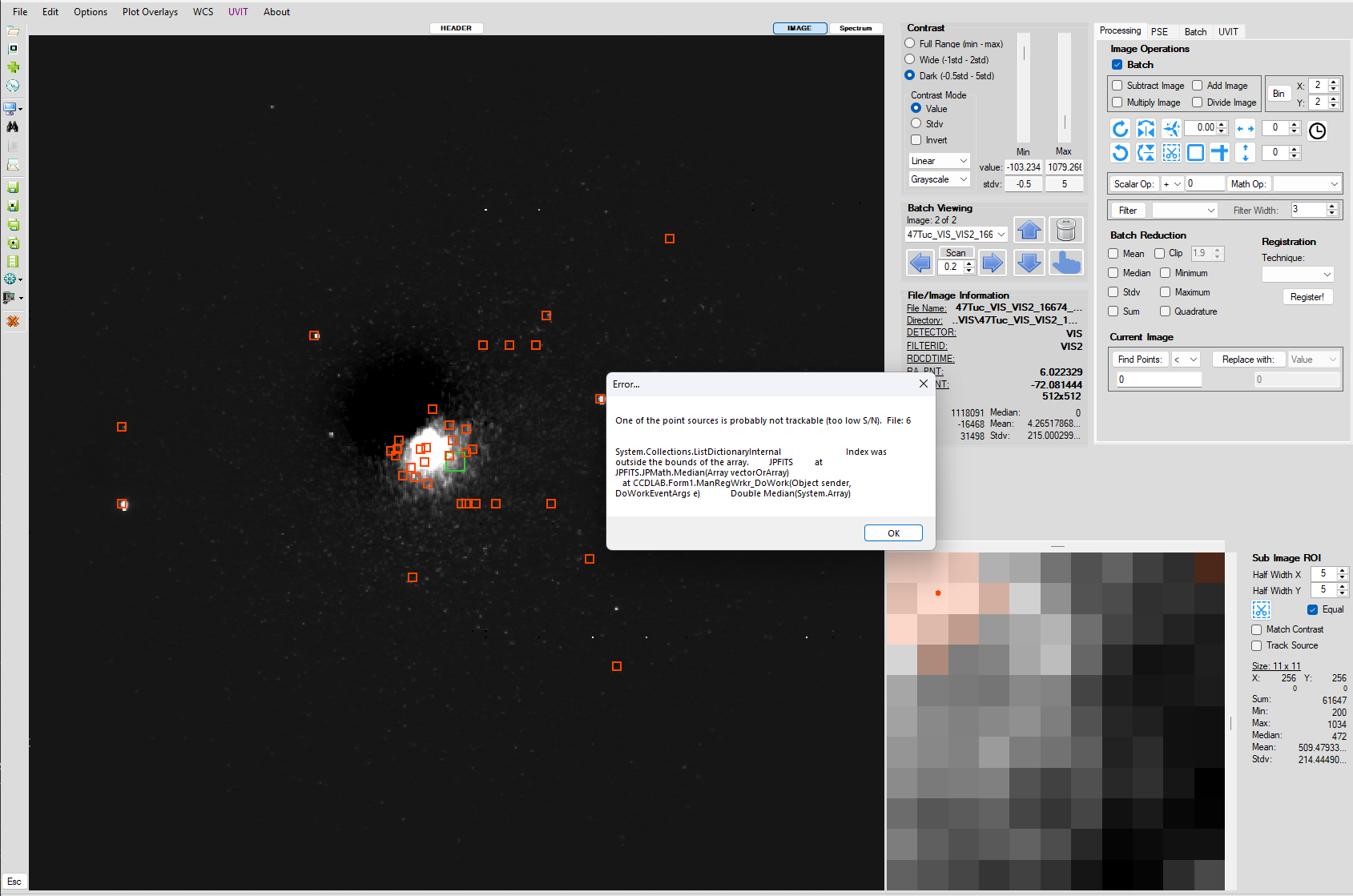}
  \includegraphics[width=\columnwidth]{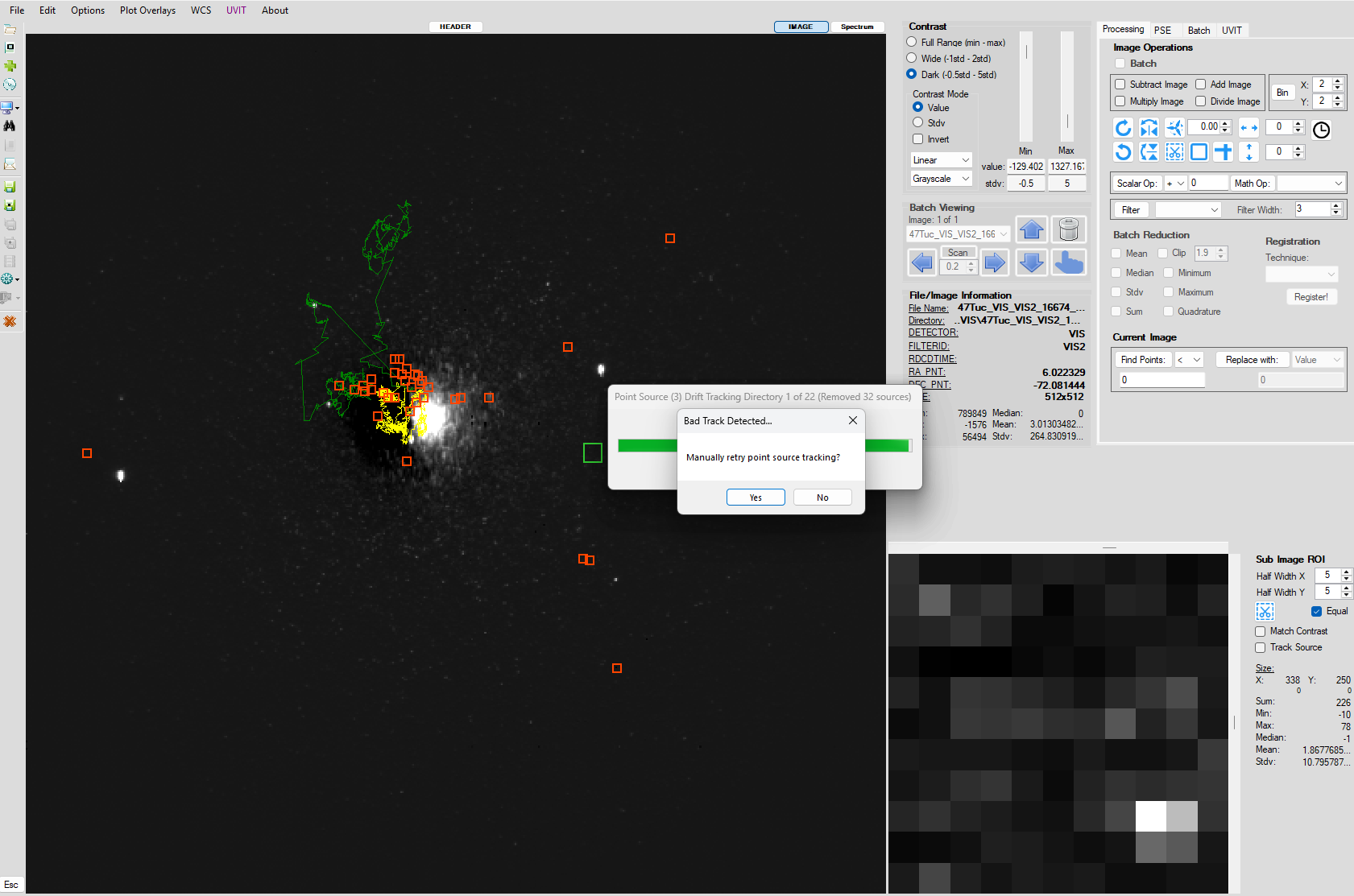}
  \caption{Error message indicating that one or more sources are not trackable (Top) and bad track detected message (Bottom).}
  \label{fig:6}
\end{figure}

\section{Manual VIS Drift Tracking} \label{sec:VIS}

In some cases, the extraction phase is interrupted because the VIS drift tracking in automated mode has not performed well due to a faint VIS field. Unlike the older version, the new version of CCDLAB displays an error message and prompts the user to manually enter point sources for tracking. Figure \ref{fig:6} shows an example of the error message (top) and the prompt to manually enter point sources (bottom).

By inspecting the tracking path of each source in the drift series, sources that have not been properly tracked can be easily identified (Figure \ref{fig:5}, bottom, yellow and green paths). The user can then manually select bright point sources, excluding those that were not properly tracked.

\begin{figure}[h]
  \includegraphics[width=\columnwidth]{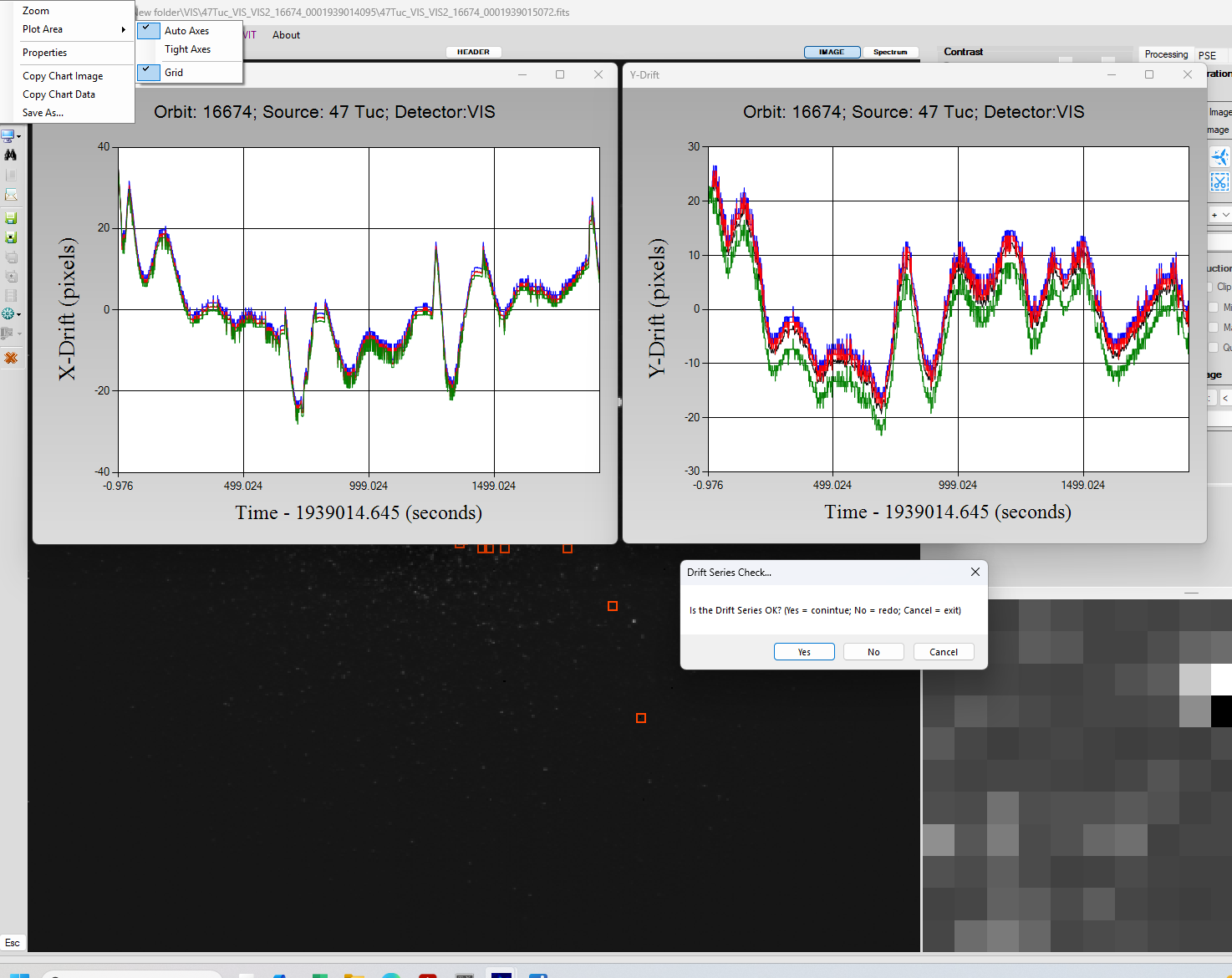}
  \includegraphics[width=\columnwidth]{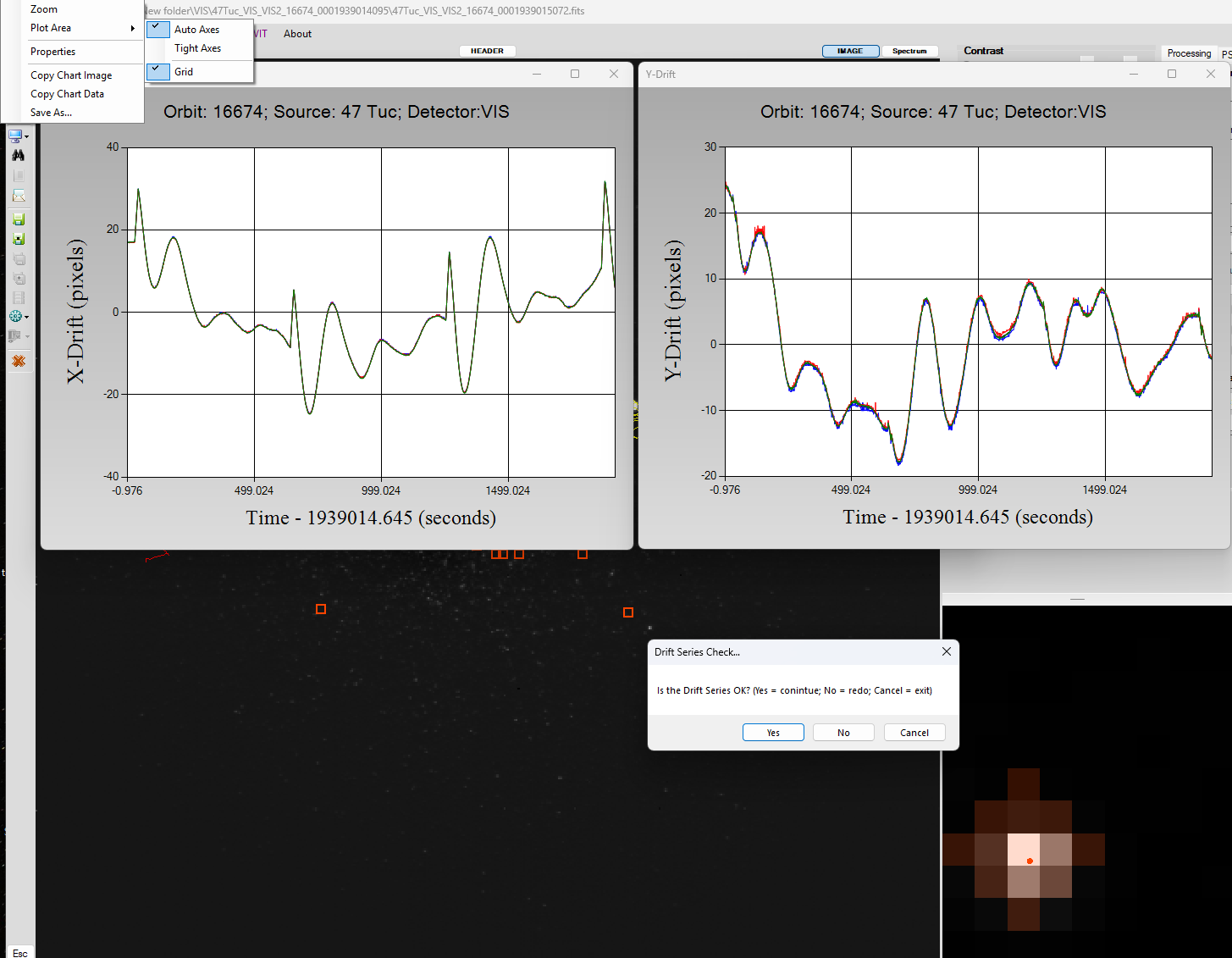}
  \caption{Drift series in x- and y-axes, automatic (Top) and manual (Bottom) point source selection.}
  \label{fig:7}
\end{figure}

Once the point sources are selected and tracking is completed, the drift series in the x- and y-axes is displayed. The plots represent multiple source drift series overlaid together. Figure \ref{fig:7} (top) shows the automated drift series in the x- and y-axes, while the bottom shows the same field drift series with manual point selection.

After the manual selection of point sources is complete and the orbit-wise images are examined, the user can proceed with the registration process as described in Section \ref{sec:datared}. The tutorial on manual VIS tracking can be found at \url{https://www.youtube.com/watch?v=RHi3j-XxYy4}. 

\section{Merging Multiple Observations of Level 1 Files of the Same Target} \label{sec:ME}

There may be multiple Level 1 files of the same target that were observed at different times or as part of different proposals. This section provides a brief description of how to process and merge these different Level 1 files into a single final science product. 

First, place the Level 1 files in a conveniently named folder and extract images individually from each Level 1 zip file, ensuring they are placed in their respective subfolders. Follow the steps in Section \ref{sec:datared}, including the registration process. Next, run the general registration on the parent folder that contains the individually extracted Level 1 files. This step should be performed manually, as image shifts may cause errors if run automatically. 

\begin{figure}[H]
  \includegraphics[width=\columnwidth]{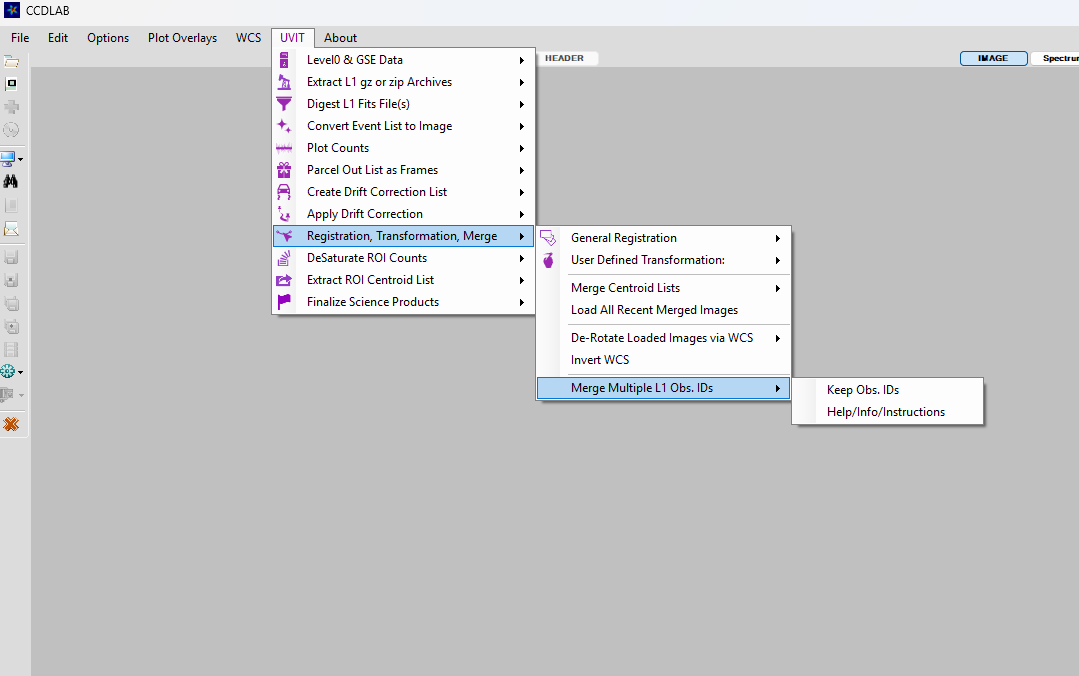}
  \caption{Merging multiple Level 1 observation IDs.}
  \label{fig:8}
\end{figure}

The next step is to double-click on ``Merge Multiple L1 Obs. IDs" under the UVIT menu: ``Registration, Transformation, Merge" and select the parent folder (see Figure \ref{fig:8}). This process moves all orbits into a single folder. To merge, double-click on ``Merge Centroid List" under the same menu and navigate to the parent folder. Finally, perform WCS correction, de-rotation, and finalization of the merged image as described in Section \ref{Sec:WCS}.

Step-by-step instructions on merging multiple observations of Level 1 files for the same target can be found in CCDLAB under ``Help/Info/Instructions" (Figure \ref{fig:8}). The tutorial is available at \url{https://www.youtube.com/watch?v=l5sDtTV_Hdc}.

This work was supported by the Canadian Space Agency through its partnership with the Indian Space Research Organization’s ASTROSAT/UVIT mission.

\end{document}